\documentclass[12pt]{iopart}

\usepackage{graphicx}

\usepackage[]{cite}

\expandafter\let\csname equation*\endcsname\relax
\expandafter\let\csname endequation*\endcsname\relax

\usepackage{amsmath}
\usepackage[]{amssymb}
\usepackage[]{mathrsfs}
\usepackage[]{eucal}
\usepackage[]{tensor}
\usepackage[]{amsthm}
\usepackage[]{bm}

\newcommand{\pf}[1]{\mathbf{#1}}
\newcommand{\dd}{\partial}
\newcommand{\hdg}{\star} 
\newcommand{\df}{\mathrm{d}}
\newcommand{\w}{\wedge}
\newcommand{\veps}{\bm{\epsilon}}
\newcommand{\Lie}{\pounds}
\newcommand{\nab}[1]{\nabla_{\!#1}}

\newcommand{\qqd}{\ , \quad}
\newcommand{\bc}{\begin{center}}
\newcommand{\ec}{\end{center}}
\newcommand{\be}{\begin{equation}}
\newcommand{\ee}{\end{equation}}

\newcommand{\F}{\pf{F}}
\newcommand{\A}{\pf{A}}

\newcommand{\FF}{\mathcal{F}}
\newcommand{\GG}{\mathcal{G}}
\newcommand{\LL}{\mathscr{L}}

\newcommand{\dis}{\mathscr{D}}
\newcommand{\balpha}{\bm{\alpha}}
\newcommand{\bbeta}{\bm{\beta}}

\newcommand{\defeq}{\mathrel{\mathop:}=}

\usepackage{dsfont}
\newcommand{\nn}{\mathds{N}}

\newcommand{\rr}{\mathds{R}}

\usepackage[unicode]{hyperref}
\usepackage{xcolor}
\definecolor{pastgreen}{HTML}{669900}
\definecolor{pastblue}{HTML}{336699}
\definecolor{pastred}{HTML}{990000}
\definecolor{linkcol}{HTML}{663333}
\hypersetup{colorlinks,linkcolor={pastblue},citecolor={pastblue},urlcolor={pastblue}}

\theoremstyle{plain} \newtheorem{tm}{Theorem}[section]
\theoremstyle{plain} \newtheorem{lm}[tm]{Lemma}
\newcommand{\btm}{\begin{tm}}
\newcommand{\etm}{\end{tm}}
\newcommand{\blm}{\begin{lm}}
\newcommand{\elm}{\end{lm}}

\begin{document}

\begin{flushright}
\texttt{ZTF-EP-23-01}
\end{flushright}

\title[Generalizations and challenges \dots]{Generalizations and challenges for the spacetime block-diagonalization}

\author{Ana Bokuli\'c and Ivica Smoli\'c}
\address{Department of Physics, Faculty of Science, University of Zagreb, Bijeni\v cka cesta 32, 10000 Zagreb, Croatia}
\eads{\mailto{abokulic@phy.hr}, \mailto{ismolic@phy.hr}}

\date{\today}

\begin{abstract}
Discovery that gravitational field equations may coerce the spacetime metric with isometries to attain a block-diagonal form compatible with these isometries, was one of the gems built into the corpus of black hole uniqueness theorems. We revisit the geometric background of a block-diagonal metric with isometries, foliation defined by Killing vector fields and the corresponding Godbillon--Vey characteristic class. Furthermore, we analyse sufficient conditions for various matter sources, including scalar, nonlinear electromagnetic and Proca fields, that imply the isometry-compatible block-diagonal form of the metric. Finally, we generalize the theorem on the absence of null electromagnetic fields in static spacetimes to an arbitrary number of spacetime dimensions, wide class of gravitational field equations and nonlinear electromagnetic fields.
\end{abstract}

\vspace{2pc}


\section{Introduction} 

Spacetimes with symmetries are certainly a mathematical idealization, nevertheless an important one, as we expect them to be a good approximation of near-equilibrium states of gravitating physical fields. Also, this reduction allows us to gain precious insight into the properties of solutions of Einstein's gravitational fields equations, most celebrated being various uniqueness and no-hair theorems. An important additional simplification of the spacetime metric ansatz is a possibility to write it (at least in some coordinate chart) in a block-diagonal form, which comes with a formal question: Under which assumptions this is not a mere convenience but a necessity? Grant and Vickers \cite{GV09} have proved that each point of an \emph{analytic} 4-manifold with an \emph{analytic} metric (of any signature) has a neighbourhood with a coordinate system in which the metric has a ``$2\times 2$'' block-diagonal form. As an aside, lower-dimensional cases admit even stronger result, local diagonalization of a \emph{smooth} metric on a \emph{smooth} manifold (again, cf.~\cite{GV09} and references therein), whereas the local metric diagonalizability in higher-dimensional cases is limited by the algebraic constraints on the Weyl tensor and its derivatives \cite{Tod92}. However, our bar here is set slightly higher, as one would like to use the coordinates adapted to the Killing vector fields and simultaneously have metric components in a block-diagonal form. To translate the dilemma into geometric language, we are looking at distributions defined by Killing vector fields and their integrability, governed by the Frobenius' theorem.

\smallskip

The problem has been essentially resolved at least for stationary electrovacuum solutions of Einstein--Maxwell field equations \cite{C73,SW93}, as one of the crucial building blocks in the architecture of black hole uniqueness theorems \cite{Heusler,CCH12}. For example, stationary axially symmetric solution of the Einstein--Maxwell's field equations on a Lorentzian 4-manifold, with the corresponding commuting Killing vector fields $k = \dd/\dd t$ and $m = \dd/\dd\varphi$, will be necessarily circular (foliated by 2-surfaces to which $k^a$ and $m^a$ are orthogonal) under a fairly mild assumptions, such as nonempty axis of axial symmetry. On the other hand, the question still has to be more thoroughly investigated for numerous theories of gravitating matter fields. Here we have, among others, nonlinear extensions of classical electrodynamics \cite{Plebanski70,Sorokin21} and various scalar fields \cite{HR15,Kobayashi19} (non-)minimally coupled to electromagnetic field and gravitation. Furthermore, once that the block-diagonal form of the metric has been established, apart from unsurprising simplifications in separation of field equations, we have some interesting additional constraints, such as the absence of null electromagnetic field in a static spacetime. Again, this result calls for a generalization to nonlinear electromagnetic fields. 

\smallskip

We shall revisit the problem of block-diagonalizability of the spacetime metric with isometries, look more carefully into the underlying geometry, generalize earlier pertaining results for various additional types of matter and discuss open questions that still have to be resolved. The paper is organized as follows. After the introduction of basic definitions and clarification of several technical details in section 2, in section 3 we analyse foliations associated with the Killing vector fields and their Godbillon--Vey characteristic class. In section 4 we present a brief summary of well-know results about integrability of distributions, while in section 5 we prove several generalizations in theories with scalar, electromagnetic and Proca fields. In section 6 we give several remarks about the integrability problem in the presence of scalar field non-minimally coupled to gravitation. In section 7 we briefly discuss new avenues for generalizations beyond general relativity, with particular focus on Carter's classification of tensors. In section 8 we give a generalization of the theorem on the absence of null electromagnetic fields in a static spacetime. Finally, in section 9 we comment on the most important open questions for this particular niche of theoretical gravitational physics. In the appendix A we summarize frequently used identities for differential forms, while in appendix B we clarify some technicalities related to Frobenius' flat chart in the presence of black hole horizons.

\section{Basic setup} 

We consider a smooth, orientable $m$-manifold $M$ with smooth Lorentzian metric $g_{ab}$, admitting a set of smooth Killing vector fields $\{ K^a_{(1)}$, \dots, $K^a_{(n)} \}$, linearly independent on a nonempty open subset $N \subseteq M$. In order to set aside trivial cases, throughout the paper we assume that $m \ge 3$ and $n \ge 1$. Points of the set $M - N$ contain, for example, axis of axial symmetry and the Killing horizon's bifurcation surfaces. Formally, we are looking at a distribution $\dis$, a smooth rank-$n$ subbundle of the tangent bundle $TN$, spanned by the $n$ linearly independent vector fields $K^a_{(1)}$, \dots, $K^a_{(n)}$. To each of these Killing vector fields metric associates a 1-form $K^{(i)}_a \defeq g_{ab} K^b_{(i)}$, also denoted by the bold letter $\pf{K}^{(i)}$. Set of 1-forms $\{ \pf{K}^{(1)}$, \dots, $\pf{K}^{(n)} \}$ defines another distribution $\dis^\perp$, smooth $(m-n)$-subbundle of the tangent bundle $TN$, given at each point $p \in M$ by $\dis^\perp|_p \defeq \mathrm{Ker}\,\pf{K}^{(1)}|_p \, \cap \dots \cap \, \mathrm{Ker}\,\pf{K}^{(n)}|_p$. In other words, $X^a \in \dis^\perp|_p$ if and only if $K^{(i)}_a X^a = 0$ at point $p \in M$ for all $i \in \{1,\dots,n\}$. It is convenient to introduce abbreviations for products of these Killing vector fields, $\kappa_{ij} \defeq g_{ab} K^a_{(i)} K^b_{(j)}$, and the set of their zeros $\mathscr{Z} \defeq \{ p\in M \mid (\exists\,i): K_{(i)}^a|_p = 0 \}$. Particular focus in the analysis that follows will be on the $n$-form $\balpha \defeq \pf{K}^{(1)} \w \dots \w \pf{K}^{(n)}$. Note that by definition $\balpha \ne 0$ on the set $N$.

\smallskip

Now, a basic geometric question for any distribution is its integrability, existence of nonempty immersed submanifolds, whose tangent space at each point coincides with the given distribution. Criterion is given by the Frobenius' theorem\footnote{Historically more accurate name should be Jacobi--Clebsch--Dehana--Frobenius' theorem \cite{Hawkins05}.} \cite{LeeSmooth}: $\dis$ is integrable if and only if it is involutive, that is if $[K_{(i)},K_{(j)}]^a \in \dis$ for all $i,j \in \{1,\dots,n\}$, while $\dis^\perp$ is integrable if and only if $\balpha \w \df\pf{K}^{(i)} = 0$ for all $i \in \{1,\dots,n\}$. For concreteness we emphasize two most commonly used classes of spacetimes with symmetries. We say that an $m$-dimensional spacetime is \emph{static} if it is stationary, with the corresponding Killing vector field $k^a$, which satisfies Frobenius' condition $\pf{k} \w \df\pf{k} = 0$. Furthermore, we say that a $4$-dimensional spacetime is \emph{circular} if it is stationary and axially symmetric, with the corresponding mutually commuting Killing vector fields $k^a$ and $m^a$, which satisfy Frobenius' conditions $\pf{k} \w \pf{m} \w \df\pf{k} = 0$ and $\pf{k} \w \pf{m} \w \df\pf{m} = 0$.

\smallskip

If both distributions $\dis$ and $\dis^\perp$ are integrable and $T_p M = \dis|_p \oplus \dis^\perp|_p$ at each point $p$ of some open subset of $N$, then this open set is covered by local coordinate charts of the form $(U;z^1,\dots,z^n,y^{n+1},\dots,y^m)$, such that $(z^1,\dots,z^n)$ are the coordinates for the integral manifold of $\dis$ and $(y^{n+1},\dots,y^m)$ are the coordinates for the integral manifold of $\dis^\perp$. Consequently, components of the spacetime metric in any of these coordinate charts attain a form of a block-diagonal matrix, as $g(\dd/\dd z^i, \dd/\dd y^j) = 0$. This picture, however, breakes down at each point $q$ where there is a nonzero vector $\ell^a \in \dis|_q \cap \dis^\perp|_q$. Then, by definition, $\ell^a$ is simultaneously a linear combination of Killing vectors $K^a_{(i)}$ and orthogonal to all of them. It immediately follows that $\ell^a$ is null, $\ell^a \ell_a = 0$. Existence of such vector may be characterized via so-called ``area function'' \cite{Carter69,CC08},
\be\label{eq:W}
\mathscr{W} \defeq {\hdg(\balpha \w {\hdg\balpha})} = -n! K^{a_1}_{(1)} \dots K^{a_n}_{(n)} K^{(1)}_{[a_1} \dots K^{(n)}_{a_n]} = -\det(\kappa_{ij}) \, .
\ee
Namely, existence of vector $\ell^a$ is a question of the existence of a linear combination $\ell^a = c^i K^a_{(i)}$ with real numbers $\{c^1,\dots,c^n\}$, not all of which are zero, and such that $g_{ab} \ell^a K^b_{(i)} = 0$, that is $\kappa_{ij} c^j = 0$. This homogeneous linear system has a nontrivial solution if and only if $\mathscr{W} = 0$. Condition $\mathscr{W} = 0$ holds, for example, on Killing horizons invariant with respect to the action of all given Killing vector fields. We note in passing that the sign in (\ref{eq:W}) has been chosen so that $\mathscr{W} > 0$ in the black hole exterior domain \cite{CC08,Chrusc09}. We shall refer to the open set $O \subseteq N$ where $\mathscr{W} \ne 0$ and both $\dis$ and $\dis^\perp$ are integrable as the \emph{orthogonal-transitive domain} of the spacetime $(M,g_{ab})$. One must bear in mind that we may rely only on the existence of \emph{local} charts where metric components attain block-diagonal form and \emph{not} on the existence of such \emph{global} chart for the whole set $O$, unless some additional stronger assumptions are imposed. Some additional remarks on coordinates for so-called Frobenius' flat chart on a neighbourhood of a Killing horizon are postponed to Appendix B.

\section{Underlying foliations} 

We say that an $n$-form is \emph{decomposable} if it can be written as a wedge product of $n$ 1-forms. Any decomposable $n$-form $\balpha$, nonvanishing on the open subset $N \subseteq M$, defines a distribution $\dis^\perp$ which, by the local Frobenius' theorem, is integrable if and only if there is a 1-form $\bbeta$ such that $\df\balpha = \balpha \w \bbeta$ and, by the global Frobenius' theorem, the set of all maximal connected integral manifolds of $\dis^\perp$ forms a codimension-$n$ foliation $\mathscr{F}$ of the submanifold $N \subseteq M$. Godbillon and Vey \cite{GV71} have observed that the $(2n+1)$-form $\bbeta \w (\df\bbeta)^n$ is closed and independent of the ambiguities in the definition, $\balpha \to f\balpha$ for smooth positive function $f$ and $\bbeta \to \bbeta + \bm{\gamma}$ for any 1-form $\bm{\gamma}$ such that $\balpha \w \bm{\gamma} = 0$. Thus, we have a well-defined de Rham cohomology class, so-called (generalized) Godbillon--Vey characteristic class $\mathrm{GV}(\mathscr{F}) = [\bbeta \w (\df\bbeta)^n] \in H^{2n+1}_{\mathrm{dR}}(N)$ (see also \cite{Ghys89,Pittie}). Thurston \cite{Thurston72} has offered an intuitive interpretation of $\mathrm{GV}(\mathscr{F})$ as a measure of a ``helical wobble'' of leaves of the foliation $\mathscr{F}$. For example, if $\balpha$ is a 1-form and $X^a$ is a nonvanishing vector field normalized such that $X^a \alpha_a = 1$, then $\Lie_X \balpha = i_X \df\balpha = \bbeta - (i_X\bbeta) \, \balpha$, implying $\df\balpha = \balpha \w \Lie_X \balpha$ and $\df\Lie_X\balpha = \Lie_X\df\balpha = \balpha \w \Lie_X \Lie_X \balpha$, so that $\Lie_X \balpha \w \balpha \w \Lie_X \Lie_X \balpha \in \mathrm{GV}(\mathscr{F})$. As $\balpha \w \Lie_X \balpha$ quantifies local ``twisting'' of leaves, $\Lie_X \balpha \w \Lie_X( \balpha \w \Lie_X \balpha)$ can be interpreted as a measure of ``gyroscopic precession'' of leaves.

\smallskip

Let us turn to our concrete $n$-form $\balpha$, defined in the section 2. Using the fact that
\be
\balpha \w i_{K_{(j)}} \df\pf{K}^{(i)} = \balpha \w (\Lie_{K_{(j)}} - \df i_{K_{(j)}}) \pf{K}^{(i)} = -\balpha \w \df\kappa_{ij} \, ,
\ee
contraction of the Frobenius' condition $\balpha \w \df\pf{K}^{(i)} = 0$ with $K^a_{(j)}$ gives
\be\label{eq:hatalphaK}
\sum_{k = 1}^n \kappa_{jk} \, \widehat{\balpha}^{(k)} \w \df\pf{K}^{(i)} = (-1)^n \, \balpha \w \df\kappa_{ij} \, ,
\ee
where we have introduced an auxiliary $(n-1)$-form
\be
\widehat{\balpha}^{(i)} \defeq (-1)^{i+1} \, \pf{K}^{(1)} \w \dots \w \pf{K}^{(i-1)} \w \pf{K}^{(i+1)} \w \dots \w \pf{K}^{(n)} \, ,
\ee
obtained by removing $\pf{K}^{(i)}$ from $\balpha$ with a convenient additional sign. System of equations (\ref{eq:hatalphaK}) may be solved for $\widehat{\balpha}^{(i)} \w \df\pf{K}^{(i)}$ on the orthogonal-transitive domain $O$,
\be
\widehat{\balpha}^{(i)} \w \df\pf{K}^{(i)} = \frac{(-1)^n}{-\mathscr{W}} \, \left| \begin{array}{ccccccc} \kappa_{11} & \dots & \kappa_{1\,i-1} & \balpha \w \df\kappa_{1i} & \kappa_{1\,i+1} & \dots & \kappa_{1n} \\ \vdots & & \vdots & \vdots & \vdots & & \vdots \\ \kappa_{n1} & \dots & \kappa_{n\,i-1} & \balpha \w \df\kappa_{ni} & \kappa_{n\,i+1} & \dots & \kappa_{nn} \end{array} \right|
\ee
allowing us to evaluate
\be\label{eq:dalpha}
\df\balpha = \sum_{i=1}^n \widehat{\balpha}^{(i)} \w \df\pf{K}^{(i)} = \frac{(-1)^n}{\mathscr{W}} \, \balpha \w \df\mathscr{W} \, .
\ee
Obtained formula (\ref{eq:dalpha}) is in agreement with the Carter's equation (10) in \cite{Carter69}. From here we immediately see that
\be
\bbeta = \frac{(-1)^n}{\mathscr{W}} \, \df\mathscr{W} = (-1)^n \, \df(\ln|\mathscr{W}|)
\ee
is exact on the open set $O$, thus the corresponding Godbillon--Vey characteristic class is trivial, $\mathrm{GV}(\mathscr{F}) = 0 \in H^{2n+1}_{\mathrm{dR}}(O)$, revealing a simplicity of the underlying foliation associated with the Killing vector fields. We stress again that we have not \emph{a priori} imposed any topological constraints on the open set $O$, such as simple connectedness. A brief remark about the 1-form $\bbeta$ on the static horizon may be found at the end of the Appendix B. 

\smallskip

There is another related foliation that we can define in this setting, assuming that the Killing vector fields $K^a_{(i)}$ are pairwise commuting (cf.~remarks in the following section). Suppose that we contract $n$-form $\balpha$ with all Killing vectors except one, say (without loss of generality) $K^a_{(1)}$. This is a generalization of the 1-form which appears in the proof of the weak rigidity for circular spacetimes (cf.~Theorem 6.13 in \cite{Heusler}). Then the 1-form
\be
\pf{L} \defeq i_{K_{(2)}} \cdots i_{K_{(n)}} \balpha
\ee
is hypersurface orthogonal. Namely, we have
\be
\df\pf{L} = (-1)^{n-1} i_{K_{(2)}} \cdots i_{K_{(n)}} \df\balpha
\ee
and
\be
\pf{L} \w \df\pf{L} = (-1)^{n-1} i_{K_{(2)}} \dots i_{K_{(n)}} \big( \balpha \w i_{K_{(2)}} \cdots i_{K_{(n)}} \df\balpha \big) \, .
\ee
Note that each term in the parenthesis is annihilated as they either contain trivial zero $\pf{K}^{(i)} \w \pf{K}^{(i)}$ or are proportional to $\balpha \w \df\pf{K}^{(i)}$, which is by assumption zero on the orthogonal-transitive domain $O$. Hence, the Frobenius' condition $\pf{L} \w \df\pf{L} = 0$ is fulfilled and contraction with $K_{(1)}^a$ leads to
\be
(i_{K_{(1)}} \pf{L}) \, \df\pf{L} = -\pf{L} \w \df(i_{K_{(1)}} \pf{L}) \, .
\ee
Since $i_{K_{(1)}} \pf{L}$ is, up to a sign, equal to the area function $\mathscr{W}$, it immediately follows that on the set $O$ we can choose the exact 1-form $\bbeta = -\df (\ln(i_{K_{(1)}} \pf{L}))$, thus the corresponding Godbillon--Vey characteristic class is again trivial.

\section{Integrability problem} 

As integrability of distributions $\dis$ and $\dis^\perp$ significantly simplifies the geometry of the problem, part of the agenda in the black hole uniqueness programme \cite{Heusler,CCH12} was to derive them from some basic assumptions, such as the field equations and boundary conditions. Involutivity of $\dis$ is a mild condition, given that a Lie bracket of Killing vector fields is again a Killing vector field and that maximal number of linearly independent Killing vector fields is finite \cite{Wald}. If an even stronger condition is met, commuting of the given Killing vector fields, $[K_{(i)},K_{(j)}]^a = 0$, then by the Theorem 9.46 in \cite{LeeSmooth} we may choose local coordinates such that $K_{(i)} = \dd/\dd z^i$ for all $i \in \{1,\dots,n\}$. Although Killing vectors generally do not satisfy this property, in some specific situations one can at least prepare commuting Killing vector fields from given ones. For example, if a Killing vector field $K^a_{(1)}$ is nonspacelike, and a Killing vector field $K^a_{(2)}$ is spacelike with compact orbits, then by the Carter's theorem \cite{Carter70} and Szabados' generalization \cite{Szabados87} we may construct a new nonspacelike Killing vector field $\widetilde{K}^a_{(1)}$, commuting with $K^a_{(2)}$, by integration of the pull-back of $K^a_{(1)}$ along the orbits of $K^a_{(2)}$. The most important application of this result is in the case of stationary, axially symmetric spacetimes with the associated Killing vector fields, timelike $k^a$ and spacelike $m^a$, for which we can, without loss of generality assume to be commuting, $[k,m]^a = 0$. A more basic question is whether a stationary solution of gravitational field equations has to be necessarily axially symmetric. The answer in general is \emph{negative}, as shown by the electrovacuum multi-black hole Majumdar--Papapetrou spacetime \cite{Papa47,Majumdar47,HH72,Myers87} and even a single-black hole spacetime \cite{RW95a,RW95b} with electromagnetic and massive vector field. On the other hand, we know that at least for asymptotically flat, analytic solution of the vacuum Einstein's field equation, by Hawking's rigidity theorem \cite{Hawking72}, its higher-dimensional generalization \cite{HIW07,HI09} and more recent generalization beyond General Relativity \cite{HIR22}, stationarity implies axial symmetry, with certain number of mutually commuting, spacelike Killing vector fields with compact orbits. An important step in elimination of the analiticity assumption was made relatively recently by Alexakis, Ionescu and Kleinerman \cite{AIK10} (review of related technical problems can be found in \cite{IK15}).

\smallskip

Integrability of the distribution $\dis^\perp$ is much less trivial. A strategy distilled through early work in 4-dimensional spacetimes \cite{KT66,Papa66,Weinstein96} and generalized to arbitrary number of dimensions \cite{IU03,CC08,Costa09}, relies on the identity (proven in the Appendix A)
\be\label{eq:alphadK}
\df{\hdg(\balpha \w \df\pf{K}^{(i)})} = 2 \, {\hdg(\balpha \w \pf{R}(K_{(i)}))} \, .
\ee
In the case when the metric is a solution of the vacuum Einstein field equation 
\be\label{eq:Einstein}
R_{ab} - \frac{1}{2} \, R g_{ab} + \Lambda g_{ab} = 0 \, ,
\ee
Ricci tensor will be proportional to the metric, $(m-2)R_{ab} = 2\Lambda g_{ab}$, implying that the $(m-n-2)$-form ${\hdg(\balpha \w \df\pf{K}^{(i)})}$ is closed for each $i$. If $m = n+2$ then ${\hdg(\balpha \w \df\pf{K}^{(i)})}$ is just a scalar, constant on each connected component of $N$. Then, given that set $M - N$ is nonempty, $\balpha \w \df\pf{K}^{(i)}$ is by continuity identically zero on each connected component of $N$ with nonempty boundary (intuitively, components of $N$ adjacent to the set of points where $\balpha$ is zero), establishing that $\dis^\perp$ is integrable. Most significantly, this has been applied to stationary, axially symmetric vacuum solutions with $m-3$ linearly independent, mutually commuting axial Killing vector fields. On the other hand, if the spacetime metric admits fewer symmetries, that is $m > n+2$, then we have to find appropriate divergence identities \cite{Heusler} which might help us to establish integrability of the distribution $\dis^\perp$.

\section{Integrability imposed by the matter fields} 

When we are considering non-vacuum Einstein field equation,
\be
R_{ab} = 8\pi T_{ab} + \frac{2\Lambda - 8\pi g^{cd} T_{cd}}{m-2} \, g_{ab} \ ,
\ee
integrability of the distribution $\dis^\perp$ boils down to vanishing of the $(n+1)$-form $\balpha \w \pf{T}(K_{(i)})$. We shall look more carefully into two most important classes of matter, described by scalar and vector fields.

\medskip

\emph{Scalar fields}. Suppose that the energy-momentum tensor of a real scalar field has a form
\be
T_{ab}^{\mathrm{(rs)}} = F(g_{cd},\phi,\nab{c}\phi,\dots) \nab{a}\phi \nab{b}\phi + G(g_{cd},\phi,\nab{c}\phi,\dots) g_{ab}
\ee
with some real functions $F$ and $G$. This covers canonical case, in which $F = 1$ and $G = -(\nab{a}\phi \nabla^a\phi)/2 - \mathscr{U}(\phi)$, with some potential $\mathscr{U}$, as well as more general k-essence theories. Then, given that scalar field $\phi$ inherits the spacetime symmetries \cite{ISm15,ISm17,FS21}, $K_{(i)}^a \nab{a} \phi = 0$ for all $i$, we immediately have $\balpha \w \pf{T}(K_{(i)}) = 0$. Complex scalar fields, with the energy-momentum tensor of the form
\be
T_{ab}^{\mathrm{(cs)}} = F(g_{cd},\phi,\phi^*,\dots) \nab{(a}\phi^* \nab{b)}\phi + G(g_{cd},\phi,\phi^*,\dots) g_{ab}
\ee
can be treated with the identical argument. Further generalization for self-gravitating scalar fields of the form $\phi : M \to S$, where $S$ is a (target) Riemannian manifold, was given by Heusler \cite{Heusler95}.

\medskip

\emph{Electromagnetic field}. We shall consider two generalizations of the Maxwell's electromagnetism:
\begin{itemize}
\item[(a)] Nonlinear electromagnetism (NLE), family of theories with the Lagrangian density $\LL(\FF,\GG)$, which is a function of two electromagnetic invariants, $\FF = F_{ab} F^{ab}$ and $\GG = F_{ab}\,{\hdg F}^{ab}$, either $\LL(\FF,\GG)$ in $m = 4$ or, as $\GG$ is scalar only in $m = 4$, $\LL(\FF)$ in $m \ne 4$;
\item[(b)] Whenever the spacetime dimension $m$ is odd, we can add the gauge Chern--Simons (gCS) term, $\mu \, \pf{A} \w \F^{(m-1)/2}$, with some coupling constant $\mu$, to the electromagnetic Lagrangian.
\end{itemize}

NLE theories have been treated since the dawn of the quantum field theory (the two prominent examples being Born--Infeld's and Euler--Heisenberg's Lagrangians \cite{Dunne04} as well as, more recent, ModMax theory \cite{BLST20}) and for the past several decades have caught a considerable attention of the gravitational community for their role in physics of compact astrophysical objects with strong magnetic fields and as candidates for regularization of the spacetime singularities \cite{Bronnikov00,GSB05,BJS22b}. Chern--Simons terms have geometric origin, as a secondary characteristic class, and appear in several places in physics, just to name few, related to quantum field theory anomalies, quantum Hall effect and topological quantum field theory. Here we assume that the electromagnetic part of the Lagrangian is, generally, a sum of NLE and gCS terms,
\be
\mathbf{L}^{\mathrm{(em)}} = \LL \veps + \mu \, \pf{A} \w \F^{(m-1)/2} \, .
\ee
Corresponding source-free generalized Maxwell's (gMax) equations can be written in the following form
\be\label{eq:gMax}
\df\F = 0 \qqd \df{\hdg\pf{Z}} = \frac{m+1}{2} \, \mu\,\F^{(m-1)/2} \, .
\ee
Here we have an auxiliary 2-form $\pf{Z}$ which in $m=4$ is defined as
\be
\pf{Z} \defeq -4(\dd_\FF \LL \, \F + \dd_\GG \LL \, {\hdg\F}) \, ,
\ee
while for $m \ne 4$ it is simplified to $\pf{Z} \defeq -4 \dd_\FF \LL \, \F$. The right hand side of the second gMax equation (\ref{eq:gMax}) is, by definition, zero for even $m$. The corresponding energy-momentum tensor can be written as
\be
T_{ab}^{\mathrm{(em)}} = \frac{1}{4\pi} \, (Z_{ac} \tensor{F}{_b^c} + \LL g_{ab}) \, .
\ee
Note that the presence of the gCS term in the electromagnetic Lagrangian does not alter the electromagnetic energy-momentum tensor $T_{ab}^{\mathrm{(em)}}$.

\btm
Suppose that the $m$-dimensional spacetime $(M,g_{ab})$ admits $m-2$ smooth pairwise commuting Killing vector fields $\{K_{(1)}^a,\dots,K_{(m-2)}^a\}$, with the corresponding nonempty set of zeros $\mathscr{Z} \subseteq M$. Furthermore, suppose that this spacetime contains electromagnetic 2-form $F_{ab}$ which inherits the symmetries, $\Lie_{K_{(i)}} F_{ab} = 0$ for all $i$. Then, given that $g_{ab}$ and $F_{ab}$ are solutions of the Einstein-gMax field equations defined above, it follows that $\balpha \w \pf{T}(K_{(i)}) = 0$ for all $i$ on any open set sharing a boundary with the zero set $\mathscr{Z}$.
\etm

\emph{Proof}. Our main concern here is a term in the electromagnetic energy-momentum tensor proportional to $Z_{ac} \tensor{F}{_b^c}$. As the 1-form $\zeta_a \defeq Z_{ac} \tensor{F}{_b^c} K^b_{(i)}$ satisfies
\begin{align}
{\hdg(\balpha \w \bm{\zeta})} & = (-1)^n \, i_{K_{(n)}} \cdots i_{K_{(1)}} {\hdg\bm{\zeta}} \nonumber \\
 & = (-1)^{m+n} \, i_{K_{(n)}} \cdots i_{K_{(1)}} ({\hdg\pf{Z}} \w i_{K_{(i)}} \F) \, ,
\end{align}
the problem of integrability is shifted to the properties of the scalar $i_{K_{(i)}} i_{K_{(j)}} \F$ and the $(m-n-2)$-form $i_{K_{(n)}} \cdots i_{K_{(1)}} {\hdg\pf{Z}}$. In particular, if $m=n+2$, the latter differential form is also a scalar, in which case we may present a following simple argument. Given that $\F$ inherits spacetime symmetries, $\Lie_{K_{(i)}} \F = 0$ for all $i$, we have
\begin{align}
\df(i_{K_{(i)}} i_{K_{(j)}} \F) & = i_{K_{(i)}} i_{K_{(j)}} \df\F \, , \\
\df(i_{K_{(n)}} \cdots i_{K_{(1)}} {\hdg\pf{Z}}) & = (-1)^n i_{K_{(n)}} \cdots i_{K_{(1)}} \df{\hdg\pf{Z}} \, .
\end{align}
First gMax equation, $\df\F = 0$, implies that scalars $i_{K_{(i)}} i_{K_{(j)}} \F$ are locally constant. Furthermore, on any open set sharing a boundary with the zero set $\mathscr{Z}$, these scalars are in fact zero. Then, using the second gMax equation, same conclusion may be deduced for the scalar $i_{K_{(n)}} \cdots i_{K_{(1)}} {\hdg\pf{Z}}$, from where it follows that $\balpha \w \pf{T}(K_{(i)}) = 0$ for all $i$. \qed

\medskip

\emph{Proca field}. Massive vector fields are described by Proca Lagrangian
\be
\LL^{\mathrm{(P)}} = -\frac{1}{4}\,F_{ab} F^{ab} - \frac{\mu^2}{2}\,A_a A^a \, , 
\ee
where $\F = \df \A$ and $\mu$ is the mass parameter of the field. The corresponding field equations are
\be
\df\F = 0 \qqd \df{\hdg\F} = -\mu^2 {\hdg\A}
\ee
and the energy-momentum tensor can be written in a form
\be
T_{ab}^{\mathrm{(P)}} = T_{ab}^{(\mathrm{Max})} + \frac{\mu^2}{4\pi} \left( A_a A_b - \frac{1}{2}\,g_{ab} A_c A^c \right) ,
\ee
with
\be
T_{ab}^{(\mathrm{Max})} = \frac{1}{4\pi} \left( F_{ac} \tensor{F}{_b^c} - \frac{1}{4} \, g_{ab} F_{cd} F^{cd} \right).
\ee
Note that for any of the Killing vector fields $K^a_{(i)}$ we have
\be
4\pi \, {\hdg\big(\balpha \w \pf{T}^{\mathrm{(P)}}(K_{(i)})} \big) = (-1)^{m+n} i_{K_{(n)}} \cdots i_{K_{(1)}} ({\hdg\F} \w i_{K_{(i)}} \F) + (i_{K_{(i)}} \A) \mu^2 \, {\hdg(\balpha \w \A)} \, .
\ee
It is not too difficult to extend the previous theorem to Proca fields, given that we include one additional assumption.

\btm
Suppose that the $m$-dimensional spacetime $(M,g_{ab})$ admits $m-2$ smooth pairwise commuting Killing vector fields $\{K_{(1)}^a,\dots,K_{(m-2)}^a\}$, with the corresponding nonempty set of zeros $\mathscr{Z} \subseteq M$. Furthermore, suppose that this spacetime contains Proca 1-form $\A$ which inherits the spacetime symmetries, $\Lie_{K_{(i)}} \A = 0$ for all $i$, and is parallel to all the Killing vector fields in sense that $\balpha \w \A = 0$. Then, given that $g_{ab}$ and $F_{ab}$ are solutions of the Einstein--Proca field equations defined above, it follows that $\balpha \w \pf{T}(K_{(i)}) = 0$ for all $i$ on any open set sharing a boundary with the zero set $\mathscr{Z}$.
\etm

\emph{Proof}. The major difference with respect to the electromagnetic case is that here we have
\begin{align}
\df (i_{K_{(n)}} \cdots i_{K_{(1)}} {\hdg\F}) & = (-1)^n \, i_{K_{(n)}} \cdots i_{K_{(1)}} \df {\hdg\F} \nonumber \\
 & = (-1)^{n+1} \mu^2 \, i_{K_{(n)}} \cdots i_{K_{(1)}} {\hdg\A} \nonumber \\
 & = -\mu^2 \, {\hdg(\balpha \w \A)} \, .
\end{align}
Thus, using the assumption $\balpha \w \A = 0$ and repeating the same reasoning as in the electromagnetic case, the claim follows. \qed

\medskip

\emph{Multifield cases}. If we combine scalar and vector fields which do not interact the arguments from above can be repeated verbatim. A nontrivial challenge comes with the interacting fields. We shall consider two types of theories with scalars and electromagnetic fields. First, if one couples nonlinear electromagnetic field to the complex scalar field,
\be
\LL^{(\phi,\mathrm{em})} = \LL(\FF,\GG) - (\mathcal{D}_a \phi)^*(\mathcal{D}^a \phi) - \mathscr{U}(\phi^*\phi) \, ,
\ee
with the covariant gauge derivative $\mathcal{D}_a = \nab{a} + iqA_a$, the corresponding NLE Maxwell's equations attain the form
\be
\df\F = 0 \qqd \df{\hdg\pf{Z}} = 4\pi{\hdg\pf{J}} \, ,
\ee
with the current 1-form
\be
J_a = \frac{iq}{4\pi} \left( \phi^* \mathcal{D}_a \phi - \phi (\mathcal{D}_a \phi)^* \right) .
\ee
The total energy-momentum tensor in this theory is given by
\be
4\pi T_{ab} = Z_{ac} \tensor{F}{_b^c} + \LL g_{ab} + 2 (\mathcal{D}_{(a}\phi)^* \mathcal{D}_{b)}\phi - \big( (\mathcal{D}_{c}\phi)^* (\mathcal{D}^c\phi) + \mathscr{U}(\phi^*\phi) \big) g_{ab} \, .
\ee
Crucial detail is that for symmetry inheriting scalar field we have a simplification $2K^b_{(i)} (\mathcal{D}_{(a}\phi)^* \mathcal{D}_{b)}\phi = -4\pi(K^b_{(i)} A_b) J_a$. This allows us to produce a straightforward generalization, given that an additional assumption is included for the current 1-form.

\btm\label{tm:Fphi}
Suppose that the $m$-dimensional spacetime $(M,g_{ab})$ admits $m-2$ smooth pairwise commuting Killing vector fields $\{K_{(1)}^a,\dots,K_{(m-2)}^a\}$, with the corresponding nonempty set of zeros $\mathscr{Z} \subseteq M$. Furthermore, suppose that this spacetime contains complex scalar field $\phi$ and electromagnetic 2-form $F_{ab}$, both of which inherit the spacetime symmetries, $\Lie_{K_{(i)}} \phi = 0$ and $\Lie_{K_{(i)}} F_{ab} = 0$ for all $i$, and $\balpha \w \pf{J} = 0$. Then, given that $g_{ab}$, $\phi$ and $F_{ab}$ are solutions of the Einstein-NLE-Maxwell-scalar field equations defined above, it follows that $\balpha \w \pf{T}(K_{(i)}) = 0$ for all $i$ on any open set sharing a boundary with the zero set $\mathscr{Z}$.
\etm

\smallskip

Next, we consider a theory containing real scalar fields nonminimally coupled to the NLE fields, the most important example being dilatons \cite{GM88}, with the matter Lagrangian density of the form
\be
\LL^{(\mathrm{dil,em})} = f(\phi)\LL(\FF,\GG) - \frac{1}{2}\,\nab{a}\phi\,\nabla^a\phi - \mathscr{U}(\phi) \, ,
\ee
where $f$ and $\mathscr{U}$ are some smooth real functions. The corresponding field equations for the electromagnetic and dilatonic field read
\begin{align}
\df\F = 0 \qqd \df{\hdg(f(\phi)\pf{Z})} & = 0 \, ,\\
\Box\phi - \mathscr{U}'(\phi) + f'(\phi)\LL(\FF,\GG) & = 0 \, ,
\end{align}
while the energy-momentum tensor may be written as
\be
4\pi T_{ab} = 4\pi f(\phi) T_{ab}^{\mathrm{(em)}} + \nab{a}\phi \nab{b}\phi - \left( \frac{1}{2}\,\nab{c}\phi\,\nabla^c\phi + \mathscr{U}(\phi) \right) g_{ab} \, .
\ee
Then, using the same reasoning as in the electromagnetic case above, we have to look at the 1-form $\zeta_a \defeq f(\phi) Z_{ac} \tensor{F}{_b^c} K^b_{(i)}$, and rely on the relation
\be
\df(i_{K_{(n)}} \cdots i_{K_{(1)}} {\hdg f(\phi)\pf{Z}}) = (-1)^n i_{K_{(n)}} \cdots i_{K_{(1)}} \df{\hdg(f(\phi)\pf{Z})} = 0 \, .
\ee
This leads to the following result.

\btm
Suppose that the $m$-dimensional spacetime $(M,g_{ab})$ admits $m-2$ smooth pairwise commuting Killing vector fields $\{K_{(1)}^a,\dots,K_{(m-2)}^a\}$, with the corresponding nonempty set of zeros $\mathscr{Z} \subseteq M$. Furthermore, suppose that this spacetime contains dilaton field $\phi$ and electromagnetic 2-form $F_{ab}$, both of which inherit the spacetime symmetries, $\Lie_{K_{(i)}} \phi = 0$ and $\Lie_{K_{(i)}} F_{ab} = 0$ for all $i$. Then, given that $g_{ab}$, $\phi$ and $F_{ab}$ are solutions of the Einstein-dilaton-Maxwell field equations defined above, it follows that $\balpha \w \pf{T}(K_{(i)}) = 0$ for all $i$ on any open set sharing a boundary with the zero set $\mathscr{Z}$.
\etm

\smallskip

As some of the recently discovered hairy black hole solutions \cite{HR15} harbour symmetry noninheriting fields, it is interesting to briefly look at what happens with the conclusions above in such a case. More concretely, let us consider the theory introduced in theorem \ref{tm:Fphi}, but assume that the scalar field breaks symmetries in a sense that $K^a_{(i)} \nab{a} \phi = i\alpha_i \phi$ with some real constants $\alpha_i \in \rr$, while the electromagnetic field inherits the symmetries and rest of the assumptions remain unaltered. This implies $2K^b_{(i)} (\mathcal{D}_{(a}\phi)^* \mathcal{D}_{b)}\phi = 4\pi ((\alpha_i/q) - K^b_{(i)} A_b) J_a$, so that conclusions of the theorem still hold (e.g.~Herdeiro--Radu black hole solution \cite{HR14} is ``forced'' to have block-diagonal metric). On the other hand, if one admits more general form of the symmetry inheritance breaking \cite{ISm15,BGS17}, the analysis needs to be thoroughly revised, which we leave for the future work.

\section{An aside on scalar field nonminimally coupled to gravitation} 

Let us turn briefly to a class of theories in which a generalization of the results from the previous section proves to be more challenging. If a real scalar field $\phi$ is nonminimally coupled to gravitation, for example, with the total Lagrangian density of the form
\be
\LL = \frac{1}{16\pi}\,R - f(\phi)R - \frac{1}{2}\,\nab{a}\phi\,\nabla^a\phi - \mathscr{U}(\phi) \, ,
\ee
then the corresponding gravitational field equation is
\be\label{eq:nonmin}
(1 - 16\pi f(\phi)) G_{ab} = 8\pi T_{ab}^{(\phi)} \, ,
\ee
with the energy-momentum tensor
\begin{align}
T_{ab}^{(\phi)} & = (1 - 2f''(\phi)) \nab{a}\phi \nab{b}\phi -2f'(\phi) \nab{a}\nab{b}\phi \nonumber \\
 & + \left( -\frac{1}{2}\,(1 - 4f''(\phi)) \nab{c}\phi \nabla^c\phi - \mathscr{U}(\phi) + 2f'(\phi) \Box\phi \right) g_{ab} \, .
\end{align}
Crucial ``problematic term'' here is the one containing second derivative $\nab{a}\nab{b}\phi$. A possible resolution relies on two following observations. First, for any Killing vector field $K^a$ we have
\be
K^a \nab{a}\nab{b}\phi = \nab{b}(K^a \nab{a}\phi) + (\nabla^a \phi) \nab{[a} K_{b]} \, .
\ee
Second, assuming that scalar field $\phi$ inherits spacetime symmetries, $\Lie_{K_{(i)}} \phi = 0$ for all $i$, and using $X^a \defeq \nabla^a \phi$, we may write
\begin{align}
{\hdg(\balpha \w i_X \df\pf{K})} & = (-1)^n \, {\hdg(i_X (\balpha \w \df\pf{K}))} \nonumber \\
 & = (-1)^{m+n+1} \, {\hdg(\balpha \w \df\pf{K})} \w \df\phi \, .
\end{align}
This allows us to extract the relation
\be
(1 - 16\pi f(\phi)) 2\,{\hdg(\balpha \w \pf{R}(K_{(i)}))} = (-1)^{m+n} \, 16\pi \, {\hdg(\balpha \w \df\pf{K}_{(i)})} f'(\phi) \, \df\phi
\ee
from the field equation (\ref{eq:nonmin}). Finally, let us assume that $m=n+2$ and denote by $U \subseteq M$ the open set on which $16\pi f(\phi) \ne 1$ and ${\hdg(\balpha \w \df\pf{K}_{(i)})} \ne 0$ for all $i$. Then, on each point of the set $U$ this relation may be rewritten as
\be
\df \ln{\hdg(\balpha \w \df\pf{K}_{(i)})} = -\df \ln(1 - 16\pi f(\phi)) \, .
\ee
As the right hand side is independent of $i$, it follows that there are real constants $\{C_1,\dots,C_{m-2}\}$ such that
\be
C_1 \, {\hdg(\balpha \w \df\pf{K}_{(1)})} = \dots = C_{m-2} \, {\hdg(\balpha \w \df\pf{K}_{(m-2)})}
\ee
on the set $U$. Unfortunately, it is not clear if these constraints are sufficient to prove integrability of any of the distributions induced by given Killing vector fields.

\section{Block-diagonalizable gravitational field equations} 

Moving away from the Einstein's field equation, we may assume that the generalized gravitational field equation has the form
\be\label{eq:ET}
\mathcal{E}_{ab} = 8\pi T_{ab} \, ,
\ee
where $T_{ab}$ is the energy-momentum tensor and $\mathcal{E}_{ab}$ is a symmetric tensor constructed from the metric $g_{ab}$, Riemann tensor $R_{abcd}$, Levi--Civita tensor $\epsilon_{a_1 \dots a_m}$, covariant derivative $\nab{a}$ and possibly even other fields, nonminimally coupled to gravitation in the action. Given that we put symmetries into a focus of discussion, one can distinguish two directions of investigations:
\begin{itemize}
\item[(a)] Assuming that all fields $\tensor{\psi}{^a^\dots_b_\dots}$ contributing to the energy-momentum tensor inherit spacetime symmetries, prove integrability of the distribution $\dis^\perp$;

\item[(b)] Assuming that $\dis$ and $\dis^\perp$ are integrable, find additional constraints on the symmetry inheritance of fields.
\end{itemize}

The essential challenge in the direction (a) is that we are losing direct link to the form $\df{\hdg{\df\pf{K}}}$, lying at the heart of the argument above, encapsulated by the identity (\ref{eq:alphadK}). One promising step forward was already paved in Carter's paper \cite{Carter69}, with a non-exhaustive classification of tensors into, what we shall refer to as, \emph{even} and \emph{odd orthogonal-transitive type tensors} (shortened as even and odd ``o-t tensors''). Suppose we form scalars by contracting given rank-$k$ tensor $\tensor{T}{^a^\dots_b_\dots}$ with $s$ Killing vectors $K^a_{(i)}$ and $k-s$ vectors from $\dis^\perp$ on the orthogonal-transitive domain $O$. If all these scalars with even (odd) $s$ are zero we say that the tensor $\tensor{T}{^a^\dots_b_\dots}$ is of even (odd) o-t type. For example, as $\bm{\alpha} \w \pf{R}(K_{(i)}) = 0$ on the domain $O$, contraction with any vector $Y_{(j)} = \dd/\dd y^j$ from $\dis^\perp$ implies $R(K_{(i)},Y_{(j)}) = 0$, thus Ricci tensor is an example of an odd o-t tensor. A more systematic exploration of the ``o-t classification'' may be performed either covariantly (cf.~supplementary material to \cite{XZSdRWY21}) or using coordinate charts $(U;z^1,\dots,z^n,y^{n+1},\dots,y^m)$ introduced in the section 2. An important lemma is that the Riemann tensor $R_{abcd}$, as well as any of its covariant derivatives $\nab{a_1} \dots \nab{a_p} R_{bcde}$, are odd o-t tensors. This allows one to prove \cite{ISm17} that tensor $\mathcal{E}_{ab}$ for the Lovelock gravity and $f(R)$ theories are also odd o-t tensors. Here we note that Bach tensor\footnote{This tensor corresponds to the squared Weyl tensor $C_{abcd} C^{abcd}$ term in the gravitational Lagrangian \cite{Schmidt84}, which was recently studied in the context of ``quadratic gravity'' \cite{SPPP18,PSPP18,PSPP20}, an extension of the Einstein--Weyl theory.} $B_{ab} = 2 \nab{c} \nab{d}\,\tensor{C}{^c_a_b^d} + R_{cd}\,\tensor{C}{^c_a_b^d}$, defined with Weyl tensor $C_{abcd}$, is yet another example of an odd o-t type tensor. In a recent letter \cite{XZSdRWY21} the authors have demonstrated how can integrability of the distribution $\dis^\perp$ be proven pertubatively, order by order, for gravitational theories with odd o-t type tensor $\mathcal{E}_{ab}$ (similar approach has been used in \cite{HIR22} for the rigidity problem). However, in this way one tacitly introduces various analyticity assumptions, which are difficult to justify from the physical point of view (in fact, as was already mentioned above, there is a strong incentive to generalize older results by ``softening'' such assumptions). 

\smallskip

In the opposite, direction (b), we have a patchwork of somewhat related results. For example, strict constraints on the symmetry inheritance of scalar fields minimally coupled to gravitation \cite{ISm17}, indicate that essentially the only symmetry noninheriting real scalar fields are of the Wyman-type \cite{ISm15,FS21} (absent in the black hole spacetimes), while the only complex scalar black hole hair is of the Herdeiro--Radu-type \cite{HR14}. Symmetry inheritance classification for electromagnetic fields \cite{MW75,Tod06,CDPS16} and their nonlinear extensions \cite{BGS17} has many gaps which yet have to be systematically explored and closed. Even more challenging, a proper understanding of the symmetry inheritance for fields nonminimally coupled to gravitation is still in its infancy (first analysis for scalar fields, via Frobenius' theorem, was presented in \cite{BS17}). 

\smallskip

To this discussion we may add one more related observation. Let us consider a theory in an $m$-dimensional spacetime, where $m = 4k - 1$ and $k \in \nn$, with a Lagrangian which, in addition to Einstein--Hilbert Lagrangian, contains gravitational Chern--Simons term (see \cite{BCDPPS11a} and references therein). In other words, we have $\mathcal{E}_{ab} = G_{ab} + \lambda C_{ab}$, where $C_{ab}$ is the generalized Cotton tensor,
\be
C^{ab} = -\frac{k}{2^{2(k-1)}} \, \epsilon^{c_1 \cdots c_{4k-2} (a} \nab{p} \left( \tensor{R}{^{b)}_{s_1}_{c_1}_{c_2}} \tensor{R}{^{s_1}_{s_2}_{c_3}_{c_4}} \cdots \tensor{R}{^{s_{2k-2}}^p_{c_{4k-3}}_{c_{4k-2}}} \right)
\ee
and $\lambda$ is the coupling constant. Now, if spacetime is static (with the corresponding Killing vector field $k^a$), then it is not too difficult to prove that $C_{ab} k^a k^b = 0$ and $C_{ab} Y^a_{(i)} Y^b_{(j)} = 0$ for any pair of vectors $Y^a_{(i)}$ and $Y^a_{(j)}$ from the orthogonal distribution. Thus, one could say that Cotton tensor is ``o-t even'' at least for static spacetimes. On the other hand, Einstein tensor $G_{ab}$, as well as the energy-momentum tensor $T_{ab}$ for various symmetry inheriting matter fields, are odd o-t tensors. This implies that for static solutions Cotton tensor and the rest of the gravitational field equation have to vanish separately,
\be
G_{ab} - 8\pi T_{ab} = 0 = C_{ab} \, .
\ee
In other words, any static solution of the Einstein--Chern--Simons field equation $G_{ab} + \lambda C_{ab} = 8\pi T_{ab}$ is necessarily a solution of the Einstein field equation $G_{ab} = 8\pi T_{ab}$, although the converse does not hold in general. One important consequence is that we may immediately generalize various uniqueness theorems, proven for static solutions of the Einstein field equation (see e.g.~\cite{BGH84,BMG88,CS01,Rogatko03,Costa09,HI12,MuAY15}), to the Einstein--Chern--Simons case. A nontrivial constraint underlying this analysis, vanishing of the Cotton tensor for the solutions of the Einstein equation, is identically satisfied at least for a class of metrics with larger isometry group, such as spherically symmetric spacetimes \cite{BCDPPS11b,BCDPPS13}. It is important to emphasize that the argument presented in this paragraph still holds if one replaces Einstein tensor $G_{ab}$ with any odd o-t tensor.

\section{Absence of null electromagnetic fields in static spacetimes} 

Carter's classification of tensors opens a venue for another generalization of a classical result for gravitating electromagnetic fields. We say that an electromagnetic field in 4-dimensional spacetime is \emph{null} if the two associated invariants vanish, $\FF = 0 = \GG$. A well-known fact from gravitational physics is that a solution of the electrovacuum Einstein--Maxwell field equations with a null electromagnetic field is inconsistent with the static spacetime. A particularly elegant proof of this result was given by Tod \cite{Tod06}, via spinorial approach. One must bear in mind that a nontrivial null electromagnetic field may exist in a circular spacetime, as shown by G\"urses \cite{Gurses77}.

\smallskip

Here we shall demonstrate how to obtain a threefold generalization: for much more general class of gravitational and electromagnetic field equations, as well as in spacetimes with dimension different from four. Dimensional generalization comes with the definitional challenge as the Hodge dual of the 2-form $\F$ is again a 2-form only in four dimensions, hence the scalar $\GG$ has to be replaced with something else in $m \ne 4$. A natural extension can be found within the generalized Petrov classification: We say that an electromagnetic field is of type N \cite{SOLA93,OP16} at point $p \in M$ if there is a null vector $\ell^a \in T_p M$ such that $i_\ell \F = 0$ and $\bm{\ell} \w \F = 0$ at this point (in the 4-dimensional case $\F$ is of N type if and only if it is null). Furthermore, we may introduce a basis $(\ell^a,n^a,s^a_{(1)},\dots,s^a_{(m-2)})$ of the tangent space $T_p M$, consisting of two null vectors $\ell^a$ and $n^a$, and spacelike vectors $s^a_{(i)}$, such that $\ell^a n_a = -1$, $g_{ab} s^a_{(i)} s^b_{(j)} = \delta_{ij}$, while all other products are zero. Associated dual basis of the cotangent space $T^*_p M$ will be denoted by $(\bm{\ell},\pf{n},\pf{s}^{(1)},\dots,\pf{s}^{(m-2)})$. Then any electromagnetic 2-form of type N can be written as
\be\label{eq:FtypeN}
\F = f_i \, \bm{\ell} \w \pf{s}^{(i)} \, ,
\ee
with some coefficients $f_i$.

\smallskip

Now we look at the gravitational field equation (\ref{eq:ET}), with the NLE energy-momentum tensor written in a convenient form \cite{BJS21,BJS22b}, 
\be\label{eq:ETem}
\mathcal{E}_{ab} = 8\pi \left( -4\dd_\FF\LL \, T_{ab}^{\mathrm{(Max)}} + \frac{1}{m}\,\Big( g^{cd}T_{cd} - \frac{m-4}{4\pi}\,\FF \LL_\FF \Big) \, g_{ab} \right)
\ee
where
\be
T_{ab}^{\mathrm{(Max)}} = \frac{1}{4\pi} \left( F_{ac} \tensor{F}{_b^c} - \frac{1}{4} \, g_{ab} \, \FF \right) . 
\ee
Note that the electromagnetic part of the action may contain gCS term as it does not contribute to the electromagnetic energy-momentum tensor and the proof does not rely on gMax equations.

\smallskip

\btm
Suppose that spacetime metric $g_{ab}$ and the electromagnetic field $F_{ab}$ are solutions of the gravitational field equation (\ref{eq:ETem}) with the odd o-t type tensor $\mathcal{E}_{ab}$. If the metric $g_{ab}$ admits a Killing vector field $k^a$, hypersurface orthogonal on the open set $O \subseteq M$, then at each point of $O$ where $\F \ne 0$, $\F$ is of type N and $\dd_\FF \LL \ne 0$, vector $k^a$ cannot be timelike.
\etm

\emph{Proof}. First we introduce the electric 1-form $\pf{E} \defeq -i_k \F$ and the magnetic $(m-3)$-form $\pf{B} \defeq i_k {\hdg\F}$, which allow us to make a decomposition
\be\label{eq:Fdecom}
-k_a k^a \, \F = \pf{k} \w \pf{E} + {\hdg(\pf{k} \w \pf{B})} \, .
\ee
Assumptions that the Killing vector field $k^a$ is hypersurface orthogonal and that the tensor $\mathcal{E}_{ab}$ is an odd o-t type give us condition $k_{[a} \mathcal{E}_{b]c} k^c = 0$. Furthermore, field equation (\ref{eq:ETem}) and decomposition (\ref{eq:Fdecom}) reduce this condition to the equation
\be\label{eq:LFEB}
\dd_\FF \LL \, \pf{B} \w \pf{E} = 0 \, .
\ee
Using the basis introduced above and decomposition (\ref{eq:FtypeN}), we can write the electric 1-form and the magnetic $(m-3)$-form as
\begin{align}
\pf{E} & = (k^a s_a^{(i)}) f_i \, \bm{\ell} - (k^a \ell_a) f_i \, \pf{s}^{(i)} \, , \\
\pf{B} & =  f_i \, {\hdg(\bm{\ell} \w \pf{s}^{(i)} \w \pf{k} )} \, .
\end{align}
Then, using $\ell^a E_a = k^a E_a = 0$ and $E^a s_a^{(i)} = -(k^a \ell_a) \, f_i$, we have
\be
{\hdg(\pf{B} \w \pf{E})} = i_E \, {\hdg\pf{B}} = (-1)^m \Big( \sum_i f_i^2 \Big) (k^a \ell_a) \, \bm{\ell} \w \pf{k} \, .
\ee
Now, assumption $\dd_\FF \LL \ne 0$ reduces (\ref{eq:LFEB}) to ${\hdg(\pf{B} \w \pf{E})} = 0$ and assumption $\F \ne 0$ implies that $\sum_i f_i^2 \ne 0$, which leads us to $(k^a \ell_a) \, \bm{\ell} \w \pf{k} = 0$, a contradiction if $k^a$ is timelike. \qed

\medskip

Once we drop the assumption $\dd_\FF \LL \ne 0$ from the theorem, counterexamples may be found among the null electromagnetic stealth fields \cite{ISm18}, closely related to so-called universal electromagnetic fields \cite{S35,OP16,OP18,HOP18,Ortaggio22}, which may exist on a static spacetime.

\section{Discussion} 

In this paper we have revisited the problem of block-diagonalization of the spacetime metric with isometries. An insight into the underlying geometry was gained, somewhat abstractly, by looking at foliations and evaluation of Godbillon--Vey characteristic class in section 3, with several additional comments on black hole horizons in appendix B. The central problem, integrability of the distribution $\dis^\perp$ and its relation to gravitational field equation, was introduced in section 4 and generalized for several theories, containing scalar, electromagnetic and Proca fields, in sections 5 and 6. A major obstacle here is encountered in theories with matter fields nonminimally coupled to gravitation, with prominent examples such as Horndeski's theory \cite{Kobayashi19}. Here we note that at least for the reduced Horndeski's theory, so-called ``viable Horndeski'' \cite{MVCF22}, problematic terms are exactly those that we have analysed in the section 6. Another impasse is encountered in theories with Yang--Mills fields \cite{HS93}, where the approach used in section 6 is not sufficient. Even worse, natural additional symmetry assumptions in the $SU(2)$ case are proven \cite{CNL02} to be inconsistent with the asymptotic flatness. Also, all previous theorems need to be generalized for various cases of symmetry nonoinheriting fields. Far wider front of open problems, mentioned in section 7, emerges once we go beyond General Relativity. To summarize, in all of these cases it would be desirable either to prove integrability of the distribution $\dis^\perp$ under some additional, not too stringent assumptions, or find a counterexample with a nonintegrable distribution $\dis^\perp$.

\smallskip

How do we cope with a metric which is not block-diagonal? Some systematics, via ``$(2+1)+1$ slicing'' in the 4-dimensional case, was proposed by Gourgoulhon and Bonazzola \cite{GB93}, for stationary, axially symmetric, but non-circular spacetimes. More recently, there is a growing interest in non-circular black hole spacetimes, both from a phenomenological point of view \cite{EH21a,EH21b,DEH22} and within some particular gravitational theories \cite{NK20}. As with any other progress from a ``spherical cow'' (or, should we say ``block-diagonal cow''), there is no doubt that this challenge will lead to a fruitful exploration.

\ack
We would like to thank the anonymous referees for numerous useful remarks, and in particular for drawing our attention to the paper \cite{GV09}. The research was supported by the Croatian Science Foundation Project No.~IP-2020-02-9614.


\appendix 

\section{Differential forms: notation and auxiliary relations} 

Differential forms are denoted either by abstract indices, e.g.~$\omega_{abc\dots}$, or a bold letter with omitted abstract indices, e.g.~$\bm{\omega}$. Volume element, $m$-form $\epsilon_{a_1 \dots a_m}$, is sometimes denoted by the bold symbol $\veps$. For any differential form $\bm{\omega}$ and $n \in \nn$ we use abbreviation
\be
\bm{\omega}^n \defeq \underbrace{\bm{\omega} \w \dots \w \bm{\omega}}_{n \ \mathrm{times}} \, .
\ee
For any symmetric tensor $S_{ab}$ and a vector $X^a$ we denote by $\pf{S}(X)$ the 1-form $S_{ab} X^b$. Hodge dual of a differential form is denoted by star,
\be
{\hdg\omega}_{a_{p+1} \dots a_m} \defeq \frac{1}{p!} \, \omega_{a_1 \dots a_p} \tensor{\epsilon}{^{a_1}^{\dots}^{a_p}_{a_{p+1}}_{\dots}_{a_m}}
\ee
and on a Lorentzian $m$-manifold we have ${{\hdg\hdg}\,\bm{\omega}} = (-1)^{p(m-p)+1} \, \bm{\omega}$. Contraction with a vector $X^a$ is denoted by
\be
(i_X \omega)_{a_2 \dots a_p} \defeq X^b \omega_{b a_2 \dots a_p} \, .
\ee
An elementary, but highly useful operation, is so-called ``flipping over the Hodge'',
\be\label{eq:flip}
i_X {\hdg\balpha} = {\hdg(\balpha \w \pf{X})} \, ,
\ee
where the ``$X$'' on the left hand side denotes the vector $X^a$, and $\pf{X}$ on the right hand side the corresponding 1-form $X_a = g_{ab} X^b$. One must be cautious about the order of forms in consecutive contractions; for example, 
\be
i_X i_Y {\hdg\bm{\omega}} = i_X {\hdg(\bm{\omega} \w \pf{Y})} = {\hdg(\bm{\omega} \w \pf{Y} \w \pf{X})}
\ee
and, more generally,
\be
i_{X_{(1)}} \cdots i_{X_{(n)}} {\hdg\bm{\omega}} = {\hdg(\bm{\omega} \w \pf{X}^{(n)} \w \dots \w \pf{X}^{(1)})} \, .
\ee

For any smooth vector fields $X^a$, $Y^a$ and a smooth Killing vector field $K^a$ we have the following identities (first of which is known as the Cartan's formula),
\begin{align}
i_X \df + \df i_X & = \Lie_X \\
\Lie_X \df & = \df \Lie_X \\
\Lie_X i_Y - i_Y \Lie_X & = i_{[X,Y]} \\
\Lie_K \hdg & = \hdg \Lie_K
\end{align}

\smallskip

Identity (\ref{eq:alphadK}) can be established using, respectfully, equation (\ref{eq:flip}), Cartan's formula, commutation properties of the Lie derivative and Ricci lemma $\df{\hdg{\df\pf{K}}} = 2{\hdg\pf{R}(K)}$,
\begin{align*}
\df{\hdg(\balpha \w \df\pf{K}^{(i)})} & = \df i_{K_{(n)}} \cdots i_{K_{(1)}} {\hdg \df\pf{K}^{(i)}} \\
 & = (-1)^n i_{K_{(n)}} \cdots i_{K_{(1)}} \df{\hdg \df\pf{K}^{(i)}} \\
 & = (-1)^n 2 i_{K_{(n)}} \cdots i_{K_{(1)}} {\hdg \pf{R}(K_{(i)})} \\
 & = 2 \, {\hdg(\balpha \w \pf{R}(K_{(i)}))} \, .
\end{align*}

\section{Several remarks on Frobenius' flat chart} 

Given a 1-form $\pf{k}$ which satisfies condition $\pf{k} \w \df\pf{k} = 0$ on some open subset $\Omega \subseteq M$, Frobenius' theorem implies that the associated distribution $\dis^\perp$ is completely integrable (see e.g.~Theorem 19.12 in \cite{LeeSmooth}): each point of the domain $\Omega$ has a local coordinate system, so-called flat chart, $(x^1,\dots,x^m)$ such that the integral manifolds of $\dis$ are (locally) hypersurfaces $x^m = \mathrm{const}$. In other words, we have $\pf{k}(\dd/\dd x^i) = 0$ for all $i \in \{1,\dots,m-1\}$, so that locally $\pf{k} = h(x^1,\dots,x^m) \, \df x^m$ with some function $h$, so-called integrating factor. In the presence of Killing horizons, construction of such coordinate system may be subtle, intertwined with the question of regular coordinate system on the horizon.

\smallskip

Let us illustrate this on the example of a static spherically symmetric spacetime, with the metric written in some Schwarzschild-like coordinates $(t,r,\theta^1,\dots,\theta^{m-2})$,
\be\label{m:sph}
\df s^2 = -f(r) \, \df t^2 + \frac{\df r^2}{f(r)} + r^2 \df\Omega^2_{m-2} \, .
\ee
Function $f : \left< 0,\infty \right> \to \rr$ is assumed to be of $C^2$ class, with finite number of zeros, the largest of which is denoted by $r_+$ (radius of the outermost horizon), and such that limits $\lim_{r\to r_+} f'(r)$ and $\lim_{r\to r_+} f''(r)$ exist. Killing vector field $k = \dd/\dd t$ becomes null at each zero $r_i$ of the function $f$, thus $r = r_i$ hypersurface is a Killing horizon. This, however, comes with the challenge as the original coordinate system will not be regular on the horizon. Construction of regular coordinates depends on the character of the horizon. Namely, Killing vector field $k^a$ along the associated Killing horizon $H[k]$ is geodesic, $k^b \nab{b} k^a = \kappa k^a$, with the surface gravity $\kappa$, constant on each connected component of the horizon, according to the zeroth law of the black hole mechanics. A well-know result is that $\kappa = \lim_{r\to r_+} f'(r)/2$. We say that the Killing horizon is nondegenrate if $\kappa \ne 0$ and degenerate if $\kappa = 0$, and each type calls for somewhat different treatment (construction of the regular coordinate system for the extremal Reissner--Nordstr\"om black hole was originaly given by Carter \cite{Carter66}, later generalized by Lake \cite{Lake79} and then revisited by \cite{LRS00, Gao03}). Here we shall focus on the nondegenerate case only. 

\smallskip

Conventional initial step is the introduction of the tortoise radial coordinate $r^*$, via $\df r^* = \df r/f(r)$, and the associated Eddington--Finkelstein null coordinates $u = t - r^*$ and $v = t + r^*$, so that the metric (\ref{m:sph}) attains the form
\be\label{eq:muv}
\df s^2 = -f(r) \, \df u \, \df v + r^2 \df\Omega^2_{m-2} \, ,
\ee
where $r$ now stands for the function $r = r(u,v)$. Here we may quantify behaviour of the tortoise coordinate $r^*$ on the horizon. Let us assume that $f(r) > 0$ on some neighbourhood $\left< r_+, r_+ + \varepsilon \right>$ and of order $O(1 - (r_+/r))$ as $r \to r_+$, that is 
\be
(\exists\,\varepsilon,C > 0)(\forall\,r \in \left< r_+, r_+ + \varepsilon \right>) : \ 0 < f(r) \le C \cdot \left( 1 - \frac{r_+}{r} \right) . \nonumber
\ee
Then, for any $r \in \left< r_+, r_+ + \varepsilon \right>$ and $\delta \in \left< 0,\varepsilon \right>$
\be
\int_{r_+ + \delta}^r \frac{\df r'}{f(r')} \ge \frac{1}{C} \int_{r_+ + \delta}^r \frac{\df r'}{1 - (r_+/r')} = \frac{1}{C} \left( r - (r_+ + \delta) + r_+ \ln(r' - r_+) \big|_{r_+ + \delta}^r \right) ,
\ee
which diverges to $+\infty$ as $\delta \to 0^+$, implying that $\lim_{r \to r_+} r^*(r) = -\infty$. For example, in the Reissner--Nordstr\"om metric we have
\be
f_{\mathrm{RN}}(r) = \left( 1 - \frac{r_-}{r} \right) \left( 1 - \frac{r_+}{r} \right) \le 1 - \frac{r_+}{r}
\ee
for $r \ge r_+$, implying that one can use constant $C = 1$ in the argument above.

\smallskip

In order to get rid of the singular factor $f$ still present in the metric (\ref{eq:muv}), one can make additional transformation to Kruskal--Szekeres coordinates, $U = -e^{-\kappa u}$ and $V = e^{\kappa v}$ (affine parameters for the integral curves of $k^a$ along the horizon), in which the metric attains the form
\be
\df s^2 = -\frac{1}{\kappa^2} \, f(r) e^{-2\kappa r^*} \, \df U \df V + r^2 \df\Omega^2_{m-2}
\ee
with $r = r(U,V)$ and $r^* = r^*(U,V)$. Now, in order to understand the behaviour of the function $f(r) e^{-2\kappa r^*}$ as we approach the horizon\footnote{Direct application of the L'H\^{o}pital's rule to the ratio $f(r)/e^{2\kappa r^*}$ is of no use as $f'(r)/(\df e^{2\kappa r^*}/\df r) = f(r) f'(r) / (2\kappa e^{2\kappa r^*})$ and $f'(r_+) = 2\kappa$.} we may write it in the form $\exp(\ln f - 2\kappa r^*)$; given that the exponent is bounded, metric component $g_{UV}$ will be nonzero on the horizon. As the derivative of the exponent, $(f'(r) - 2\kappa)/f(r)$, remains bounded and (by L'H\^{o}pital's rule) is finite in the limit $r \to r_+$, it follows that the exponent is indeed bounded on the interval $\left< r_+, r_+ + \varepsilon \right>$. Although we have constructed a regular coordinate system, 1-form $\pf{k}$ is still not in the sought form, as
\be\label{eq:k}
\pf{k} = \kappa g_{UV}(U,V) (V \df U - U \df V) \, .
\ee
In order to write 1-form $\pf{k}$ in flat chart from the Frobenius' theorem, we may change $(U,V)$ coordinates to ``polar ones'' $(R,\Phi)$ via $V = R \cos\Phi$ and $U = R \sin\Phi$. Then
\be
\pf{k} = \kappa R^2 g_{UV}(R,\Phi) \, \df\Phi \, .
\ee
Here we may add several remarks. First, horizon components are given by $\Phi = p\pi/4$ for $p \in \{0,1,2,3\}$ and as $\tan\Phi = U/V = -e^{-2\kappa t}$, it follows that $\Phi = \mathrm{const}.$ hypersurfaces (in the black hole exterior region) coincide with the $t = \mathrm{const}.$ hypersurfaces. Finally, using (\ref{eq:k}), we see that in the Kruskal--Szekeres coordinates we have
\be
\df\pf{k} = \pf{k} \w \df (-\ln (V^2 g_{UV}))
\ee
and we may choose $\bbeta = \df (-\ln (V^2 g_{UV}))$ on a set of points where $V^2 g_{UV} \ne 0$ (e.g.~on a neighbourhood of the future horizon). However, as we rely on the \emph{local} coordinate chart, conclusion about the exactness of the 1-form $\bbeta$ holds only locally.

\section*{References}

\bibliographystyle{iopnum}
\bibliography{block}

\end{document}